\documentclass{PoS}
\newcommand{\av}[1]{\langle #1 \rangle}
\title{Optimizing the basis of $B \to K^* l^+l^-$ observables and understanding its tensions}

\ShortTitle{Understanding the tensions in $B \to K^* l^+l^-$}

\author{S\'ebastien Descotes-Genon\\        Laboratoire de Physique Th\'eorique, CNRS/Univ. Paris-Sud 11 (UMR 8627), 91405 Orsay Cedex, France}
\author{\speaker{Joaquim Matias}
\\
        Grup F\'{\i}sica Te\`orica, Universitat Aut\`onoma de Barcelona, 08193 Bellaterra, Barcelona\\
        E-mail: \email{matias@ifae.es}}
\author{Javier Virto \\    Theoretische Physik 1, Naturwissenschaftlich-Technische Fakult\"at,   Universit\"at Siegen, 
D-57068 Siegen, Germany }

\abstract{
We discuss the impact of recent LHCb data on the $B \to K^* \mu^+\mu^-$  decay for New Physics searches. We use an optimised set of observables with a limited sensitivity to hadronic inputs (form factors), leading to a significant deviation from the Standard Model for the semileptonic operator $O_9$. We ascertain the robustness of our results with respect to models of power corrections, charm-loop effects, factorisation approaches, possible effects from  resonance tails at low-recoil, choice of observables and binnings. We explain some contradictory results obtained  for the chirality-flipped operator $O_9^\prime$ by different analyses by focusing on the sensitivity of observables  at low recoil afflicted by the $\psi(4160)$  and possibly other resonances for $B^+\to K^+ \mu^+\mu^-$ and illustrating the need for a finely binned analysis at large recoil (rather than a wide [1-6] bin). Finally, we present updated results including experimental correlations in $B \to K^* \mu^+ \mu^-$ observables, confirming the deviation seen with respect to the SM hypothesis (4.2$\sigma$ for large-recoil data, 3.5$\sigma$ for large- and low-recoil data).

}

\FullConference{The European Physical Society Conference on High Energy Physics -EPS-HEP2013\\
		18-24 July 2013\\
		Stockholm, Sweden}

\begin{document}

The 4-body decay $B \to K^*(\to K\pi) \mu^+\mu^-$ might play a prominent role in the search for New Physics (NP) in Flavour by providing new constraints on Wilson coefficients. Indeed, extracting the short distance information in the Wilson coefficients entering this 4-body angular  distribution with a reduced pollution from hadronic uncertainties has been the main goal of a series of works \cite{kruger,pap1,pap2,bobeth,becirevic,pap4,pap5}. Particularly interesting is the enhanced sensitivity to the Wilson coefficient of the semileptonic operators ${O}_{9,10}$ (and chirally flipped ones). They are, up to now, only weakly constrained by other processes as compared to the strong constraints on the Wilson coefficient of the electromagnetic operator $O_7$.

Our approach originally was based on the idea of generalizing the cancellation of soft form factors at the position of the zero of the forward-backward asymmetry $A_{FB}$ to all the low  dilepton invariant mass square up to 9 GeV$^2$. This led to the so-called transverse asymmetry $A_T^2$ \cite{kruger}, the first observable that exhibited an independence w.r.t. form factors at LO in an effective theory and that was called in short a "clean observable". A recent analysis \cite{camalich} pointed out, using a different approach, the particularly strong shielding properties of this observable from hadronic uncertainties beyond form factor cancellations.
Later on a second observable, called $A_T^{re}$\cite{becirevic} was proposed sharing the same independence from soft form factors at LO. Finally, a complete basis of observables\footnote{See \cite{pap5} for more general cases (lepton masses and scalar operators)}, namely
${\cal O}_i=\{P_1,P_2,P_3,P_4^\prime,P_5^\prime,P_6^\prime, A_{FB}, d\Gamma/dq^2\}$ was proposed in \cite{pap4,pap5} to describe this distribution under the criteria of excellent experimental accessibility, simplicity in the fit and maximizing cleanliness in the large-recoil region. A full prediction for those observables in the whole range of $q^2$ (large and low $K^*$ recoil) was presented in \cite{pap6}. 

In the following we present the updated result of our analysis of recent LHCb data using this basis of observables.

\section{Analysis pre-EPS 2013: $P_1,P_2,A_{FB}$}

At the Beauty 2013 conference the first data from LHCb on two of the clean observables of our basis $P_1\equiv A_T^2$ and $P_2\equiv A_T^{re}/2$ was presented \cite{expbeauty}. The main conclusions that one could extract from this data were, on one side, that the  too large error bars on $P_1$ do not allow to draw any conclusion yet on the presence of right-handed currents, and on the other side, that $P_2$ which can be considered as the evolved version of $A_{FB}$, exhibited a small tension in its second bin in agreement with $A_{FB}$. Also both observables had a zero perfectly consistent with the SM ($q_0^{2SM}=3.95 \pm 0.38$ GeV$^2$)  but with a tendency towards a higher $q^2$ position ($q_0^{2exp}=4.9 \pm 0.9$ GeV$^2$). Even if those   tiny deviations are not statistically significant an exploration of different mechanisms that could explain both effects pointed to $C_7^{NP} < 0$ (preferred also by radiative constraints) and $C_9^{NP} < 0$. Other mechanisms involving products of chirally flipped operators were also possible (see \cite{matiastalk}). Interestingly a measurement of the $P_i^\prime$ observables (specially $P_5^\prime$) would be able to test those mechanisms.

\section{Analysis post-EPS 2013: impact of the $P_i^\prime$}
During the EPS conference the new LHCb data on the primary observables $P_i^\prime$ was presented \cite{talkserra,Aaij:2013qta}.
We included this data and repeated our analysis using a $\chi^2$ frequentist approach in \cite{thepaper}, including asymmetric errors and estimating theory uncertainties for each choice of sets of Wilson coefficients affected by NP. 

We did {\bf three types of analyses} using different sets of data: a) large-recoil data (three bins) b) large+low recoil data (five bins) c) [1-6] bin analysis. At present our analysis \cite{thepaper} is the only one in the literature exploiting the full bin-by-bin information. The reason we did the three types of analyses is because each region has a different sensitivity to Wilson coefficients. As the low-recoil region turns out to be more delicate to understand from the experimental point of view (presence of resonances or their tails), it is important to perform analyses  separately of the two regions and focus on the more reliable one (large recoil). The set of observables considered were: $P_{1,2}, P_{4,5,6,8}^\prime$, $A_{FB}$ and we add to this list the radiative and dileptonic B decays:  ${\cal B}(B\to X_s\gamma) _{E_\gamma>1.6 {\rm GeV}}$, ${\cal B}(B\to X_s \mu^+\mu^-)_{[1,6 ]}$ and ${\cal B}(B_s\to \mu^+\mu^-)$, $A_I(B\to K^* \gamma)$ and the  $B\to K^* \gamma$ time-dependent CP asymmetry $S_{K^*\gamma}$. Notice that in this first set we did not include the ${\cal B}(B \to K^* \mu^+\mu^-)$ because the very large theory error bars coming from our conservative estimates of form factors preclude to get any useful information. No experimental correlations were available, so they were not included.

The main result of our analysis showed that large-recoil data favored clearly two scenarios, with a large and negative contribution to $C_9$ as a common feature. They are discussed in the following together with the general case.
\begin{itemize}
\item{\bf Scenario $C_7,C_9$}:
We found that the SM hypothesis ($C_7^{NP}=0,C_9^{NP}=0$) exhibits a pull of 4.5$\sigma$ using large-recoil data, reduced to 3.9$\sigma$ if large and low recoil data is considered. The best fit point in this scenario $C_9^{NP}=-1.6$ and $C_7^{NP}=-0.02$ when applied to the [1-6] bin  $\av{P_5'}_{[1,6]}|_{\rm bfp}=+0.16$, reduces the tension from 2.5$\sigma$ to 0.2$\sigma$. A variant (and simplified version) of this scenario given the smallness of $C_7^{NP}$ is to fix all Wilson coefficients to zero except for $C_9^{NP}$ that prefers in this case a value around $-1.5$. 

\item{\bf Scenario $C_9,C_9^\prime$}:
Even if our previous scenario reproduces well the data in 1 to 6 bins still we see a tension in the first and third bins of $P_5^\prime$. 
We found in \cite{thepaper} that a {\bf negative} $C_9^\prime$  pulls as strongly as $C_7^{NP}$ and could help in reducing the bin-by-bin tension in $P_5^\prime$ as Fig.1 shows. However the fact that the significance of $C_9^\prime$ different from zero  is only of 1$\sigma$ together with the preference for {\bf positive} $C_9^\prime$  of the first low recoil bin tend to reduce the significance of this solution.  Notice that a large positive value (as claimed in \cite{straub}) for $C_9^\prime$ would give a worse agreement with data for the third bin of $P_5^\prime$ (see Fig.1 and discussion below). In conclusion, we see a tension between large and low-recoil concerning $C_9^\prime$.

\item{\bf General scenario}: In \cite{thepaper} we also explored the scenario with all coefficients free. We found that $C_9^{NP}$ is consistent with SM only above 3$\sigma$, $C_7^{NP}$  around 2$\sigma$ and all others are consistent with zero at 1$\sigma$.
The best fit point in this scenario if only large-recoil data is taken is $(C_7^{NP}, C_9^{NP}, C_{10}^{NP}, C_7^\prime,C_9^\prime, C_{10}^\prime)=(-0.02,-1.6,+0.18,+0.005,-1.4,-0.13)$. If this point is used to compute the anomalous bin $\av{P_5'}_{[4.3,8.68]}|_{\rm bfp-large\, rec}=-0.49$ reduces the tension with  data$-0.19^{+0.16}_{-0.16}$ at 1.8 $\sigma$. It is interesting to note  that if low-recoil data is included $C_9^\prime$ flips sign at the best fit point and becomes positive and small with a value of +0.4.  

\end{itemize}

The common final pattern is for SM operators: $C_9^{NP} \sim -1.5$, $C_7^{NP}<0$ and small, $C_{10}^{NP}$ small, while  the chirally flipped operators $C_{7,10}^\prime$ are also small and $C_9^\prime <0$ (from large-recoil) or  $C_9^\prime>0$ and small (with low-recoil). So $C_{10,7\prime, 9^\prime,10^\prime}^{NP}$ are all consistent with zero within 1$\sigma$ with present data.

Finally there is a misunderstanding in the literature that we want to clarify here. It is said that our $C_9^{NP}-C_7^{NP}$ scenario is not able to explain the anomaly bin and more Wilson coefficients are needed. Indeed if we take our best fit point $C_9^{NP}=-1.6$ and $C_7^{NP}=-0.02$ at the anomaly we get $\av{P_5'}_{[4.3,8.68]}|=-0.56$ (or -0.49 at the large-recoil best fit point of our general scenario) which is still not in agreement with data, but if we take the best fit point of \cite{straub} one gets $\av{P_5'}_{[4.3,8.68]}|=-0.74$ which is substantially worst. So we also consider that other Wilson coefficients can play a role but probably  not in the direction and sizes given in \cite{straub} as this computation shows. With the present data, in the general case, we need NP contribution above 3$\sigma$ for $C_9^{NP}$, 2$\sigma$ for $C_7^{NP}$ and less than 1$\sigma$ for all others. More data is required to clarify the size and sign of all the rest of Wilson coefficients, besides $C_9^{NP}$ (and $C_7^{NP}$)  with a reasonable significance.

\subsection{Theory framework and power corrections}
We work in the framework of NLO QCD Factorization including $\alpha_s$ factorizable and non-factorizable corrections.  Full form factors are taken from  KMPW \cite{KMPW} obtained using LCSR  out of which soft form factors are extracted (see ref. \cite{pap6} for more details). Here we provide a few comments on where and how  power corrections have been already taken into account but a fully detailed analysis   on power corrections will be found in \cite{newpaper}.

\begin{itemize}
\item {\bf Factorizable power corrections}: The determination of the $q^2$ dependence of soft form factors can be done in two ways: i) using  their HQET limit expression as  in \cite{camalich} or ii) determining them  in terms of full form factors that include power corrections as, for instance, in ref \cite{pap6}. In the first case \cite{camalich}, one is forced to add by hand possible power corrections  of the type $a_F + b_F s/m_B^2$ to correct for the HQET expression used and find a consistent procedure to fix $a_F, b_F$. In the second case \cite{pap6,thepaper,newpaper}  power corrections are partly included in the uncertainties attached to the soft form factors, since they are extracted from full form factors. One can decide to go beyond these included power corrections and check to which extent the full form factors violate the heavy-quark symmetry relations in order to estimate the size of additional power corrections. This exercise was already done in \cite{pap6} but will be detailed in \cite{newpaper}. Some preliminary results \cite{newpaper} show that a) the heavy-quark symmetry relations are  fulfilled if the second approach is used, except for the form factor $A_0$ for which we assigned in all our predictions a huge error bar (see \cite{pap6}) b) 
the importance of taking into account correlations for the correct determination of $a_F$ and $b_F$.

\item {\bf Non-factorizable power corrections}: We multiply each amplitude by $(1+ c_i e^{i \phi_i})$ and take values inside $c_i \in [-0.1,0.1]$ and $\phi_i \in [-\pi,\pi]$ and 1$\sigma$ contains 66\% of values around the median. The $\pm 10\%$ variation is purely based on dimensional arguments.

\end{itemize}

\begin{figure}	
\includegraphics[height=4.8cm,width=4.9cm]{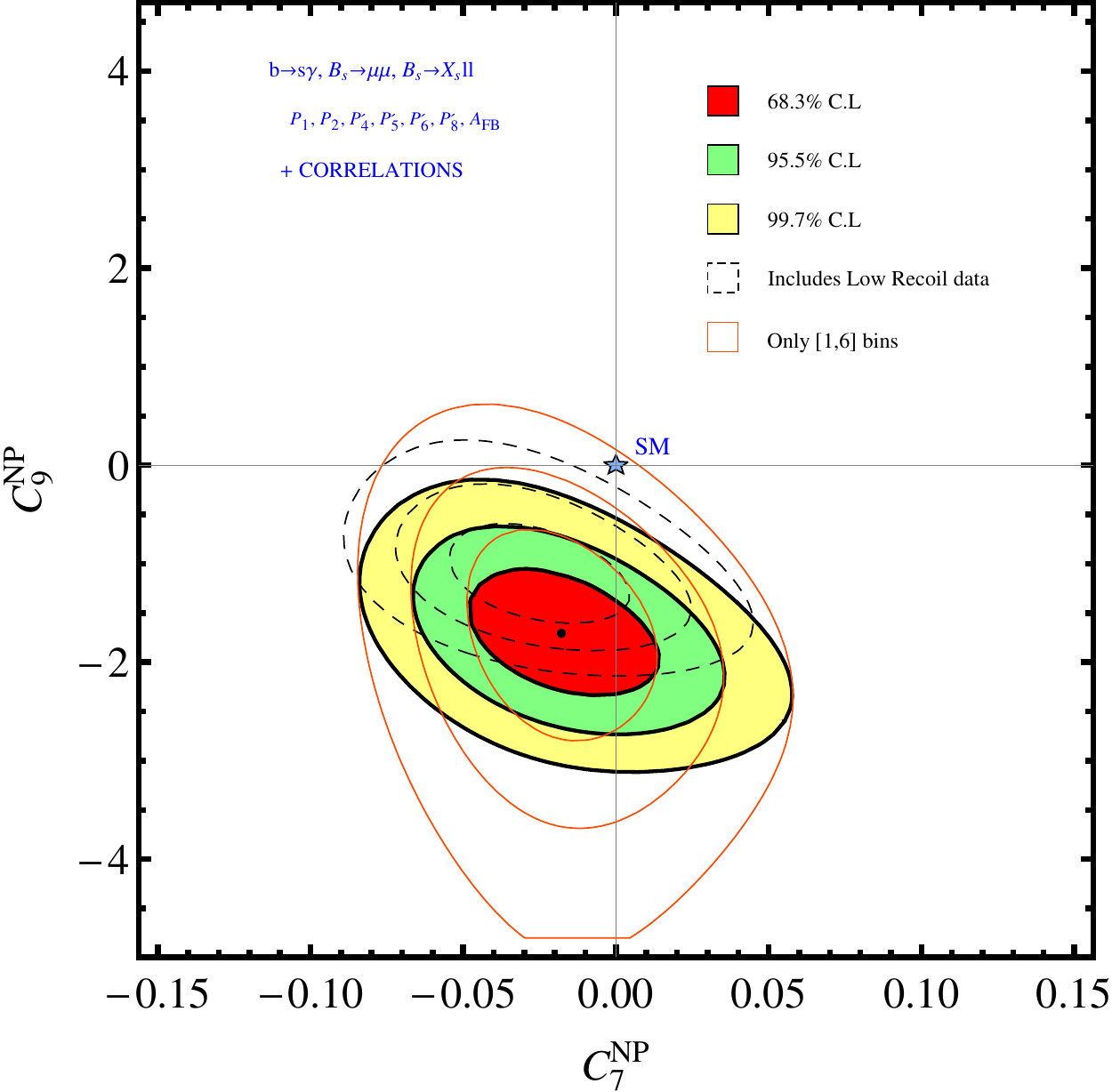}
\includegraphics[width=4.9cm,height=4.8cm]{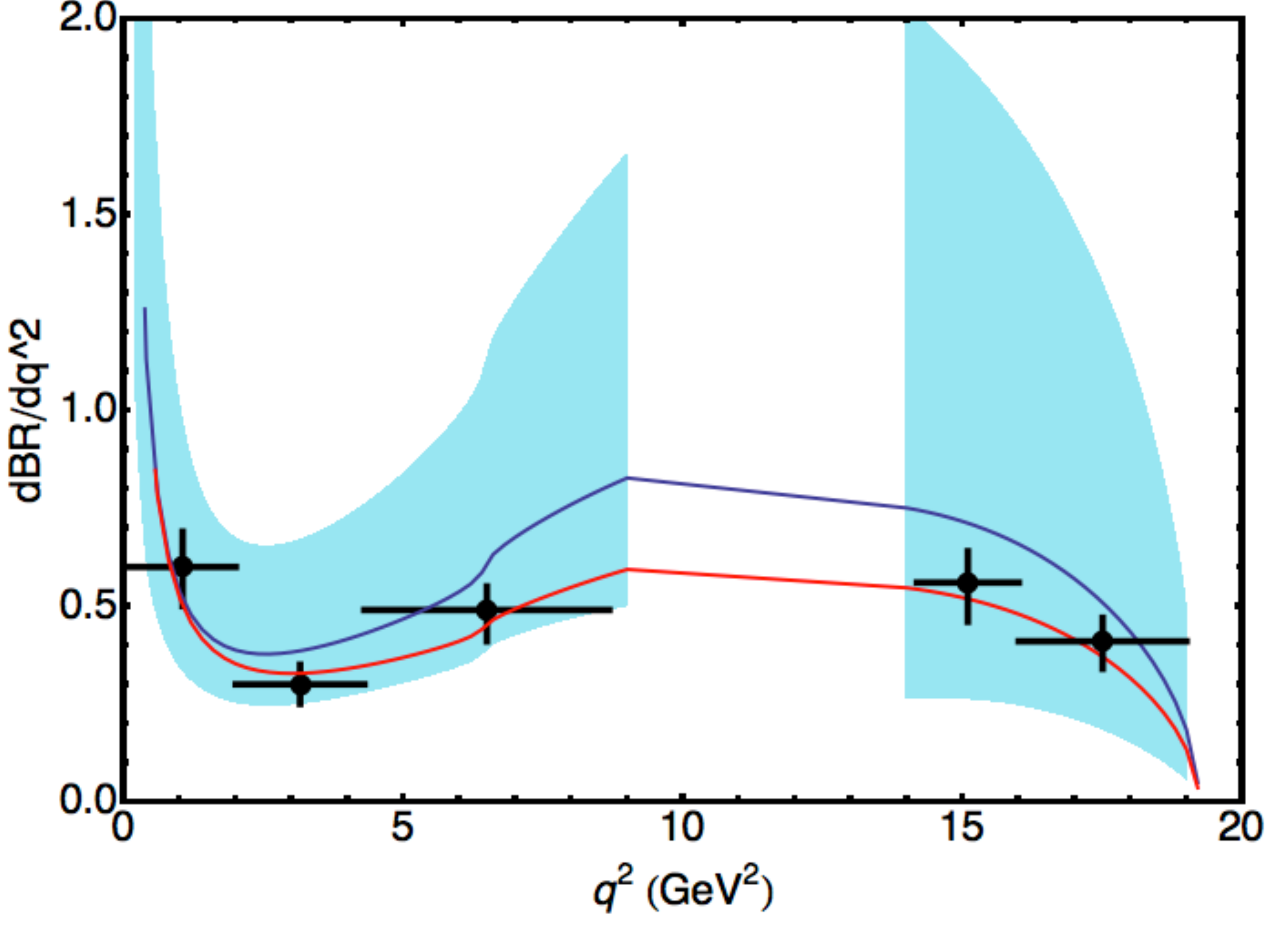}
\vspace*{0.2cm}\includegraphics[height=4.8cm,width=5cm]{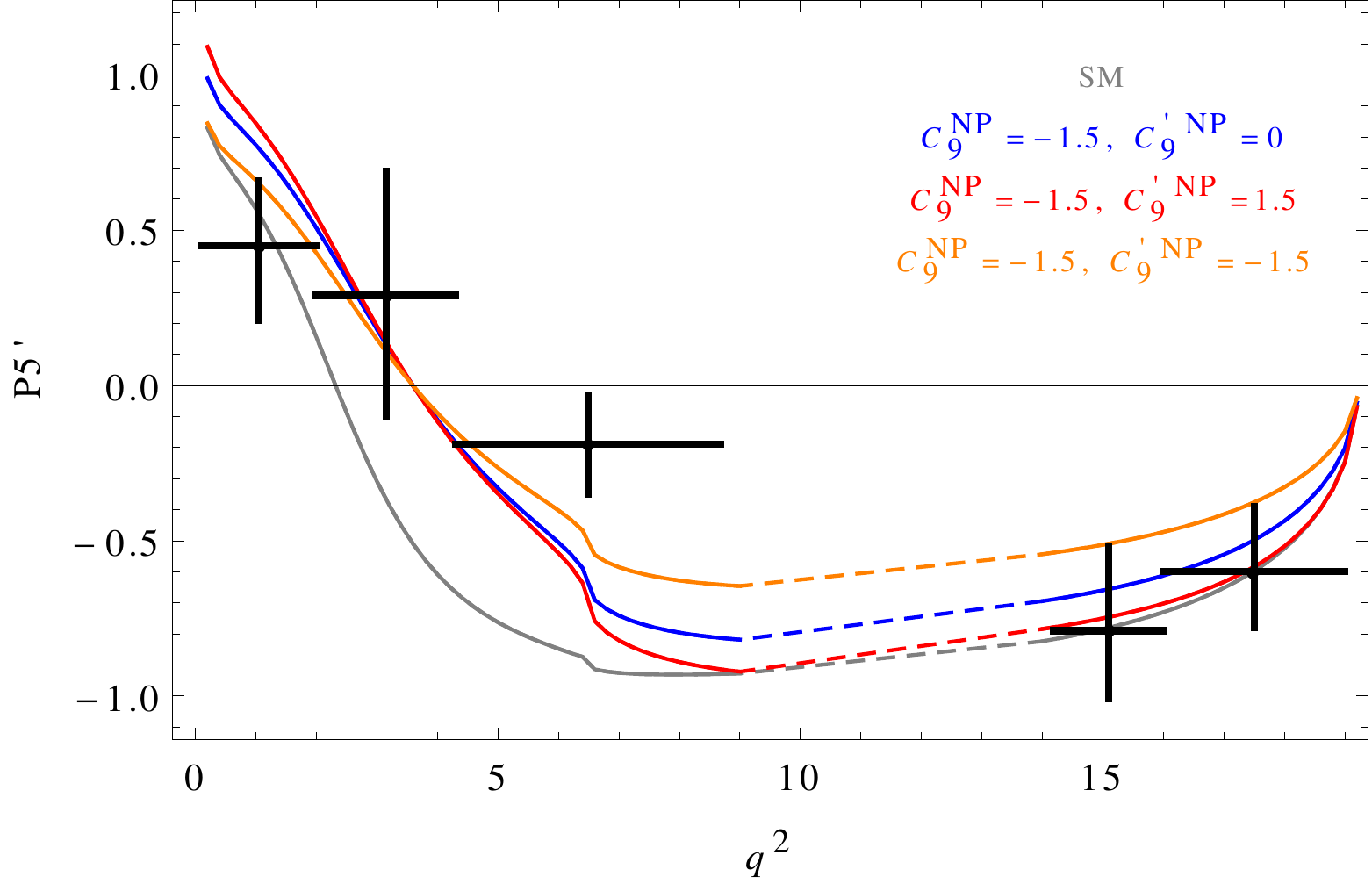}
\caption{(Left): Updated $(C_7^{NP}, C_9^{NP})$ plot including correlations. (Center): $d{\cal B}(B \to K^* \mu^+\mu^-)/dq^2$, SM (blue line) including uncertainties,  $C_9^{NP}=-1.5$ (red line). (Right): $P_5^\prime$: SM (grey line), $C_9^{NP}=-1.5$ (blue), $C_9^{NP}=-1.5$, $C_9^\prime=-1.5$ (orange),  $C_9^{NP}=-1.5$, $C_9^\prime=+1.5$ (red). $C_9^\prime<0$ improves first and third bin.}
\end{figure}

\subsection{Robustness tests}
Some of the tests  done to check the robustness of the analysis are:

\begin{itemize} 
\item {\bf Non-factorizable power corrections}: We multiplied by three the error bar associated to $\Lambda/m_b$ corrections in each observable and for each bin. This only reduces the significance  from 4.5$\sigma$ to 4$\sigma$ in the $C_7^{NP}-C_9^{NP}$ plane.

\item {\bf Charm loops}: We explored the impact of 4-quark operators (${O}_{1,2}^c$) and penguin operators $O_{3-6}$ that influence the extraction of $C_9$. We followed the prescription of KMPW\cite{KMPW} that recast the long distance effect of charm loops inside $C_9$ as an effective amplitude-dependent contribution $\Delta C_9^{\perp,\|,0}$.This contributions being positive obviously implies an enhancement of the anomaly, by increasing the difference between data and the SM prediction corrected by this long distance (see, for instance, the impact in $A_{FB}$ in KMPW\cite{KMPW}).  On the contrary, the impact of increasing the charm mass entering the perturbative contribution to $C_9^{\rm eff}$ is the opposite and would tend to reduce the anomaly mildly.

\item {\bf Influence of first low recoil bin [14.18,16]}: Due to the problems observed in $P_4^\prime$ (or $S_4$) and $P_2$ in this bin that might be affected by the tail of a charmonium resonance, we repeated the analysis removing this bin and we found that we recover the 4.5$\sigma$ significance if only the second low recoil bin together with all large-recoil bins are included in the analysis. 

\item {\bf Naive factorization} We repeated the analysis using naive factorization (with KMPW) and again we found a preference for a $C_9^{NP}$ negative albeit with much lower significance. This analysis is extremely  sensitive to the choice of form factor inputs as well as their correlations.

\item {\bf $S_i$ analysis}: We have performed a bin-by-bin analysis using the $S_i$ observables (angular observables not particularly designed to have a reduced sensitivity to form factors). Naturally the significance as expected  is lower in these observable due to their form-factor dependence. But interestingly the best-fit point in the $(C_7^{NP}, C_9^{NP})$ plane is $C_7^{NP}=-0.02$ and $C_9^{NP}=-1.76$  not far from the best fit point obtained using the clean observables. 
\end{itemize}

\subsection{Beyond Standard Model}
We have proposed a simple $Z^\prime$ model \cite{thepaper}  with couplings to left-handed quarks (with same phase as $V_{tb} V_{ts}^*$ to avoid large contributions to $\phi_s$) with flavour-changing couplings to down-type quarks and equal left and right handed couplings to charged leptons of order 0.1. The scale of the $M_Z^\prime$ is around 1-2 TeV to get $C_9^{NP}=-1.5$. An interesting implementation of our pattern for Wilson coefficients is the embedding of a $Z^\prime$ inside a specific model (3-3-1) that has been presented in \cite{florian} (see also \cite{florian2}), however in this case the scale moves towards a $M_{Z}^\prime \sim 7$ TeV due to the couplings to charged leptons of this model. Other proposals to find models that could explain the anomaly are discussed in \cite{burasmodel,diptimoy, straub}.  Finally, other well motivated models, like MSSM, warped extra dimensions and partial compositeness seems to be in trouble to reproduce the observed pattern.

\subsection{The low-recoil problem and the sign of $C_9^\prime$}

In ref. \cite{straub} an analysis using the $S_i$ observables was performed. The computation is based on naive factorization adding also non-factorizable QCDF corrections. Unfortunately this type of analysis, contrary to the $P_i$ analysis, depends crucially on the error bars of the form factor parametrization used. The authors of ref \cite{straub} use a LCSR form factor determination \cite{buras} that exhibits smaller  error bars  than KMPW. Remarkably, their results are in quite good agreement with ours, namely a large negative contribution to $C_9$, albeit smaller than in our case due to the smaller sensitivity to NP of the form factors-dependent observables  $S_i$.  One can also notice that their best-fit point  is inside our 1$\sigma$ range of our general case (once compared at the same scale). The analysis of ref.\cite{straub}  obtains also  that $C_9^\prime$ is positive and of similar size than $C_9^{NP}$ (this argument comes from the low-recoil region of $B^+ \to K^+\mu^+\mu^-$ that it is known to be affected by  the $\psi$(4160) resonance\cite{lhcbpaper}). This enters in part in contradiction with our findings that $C_9^\prime$ is 
 either negative (if only large-recoil is considered) or positive but small (if all low-recoil is included). An explanation of this difference is that ref.\cite{straub} proceeds only with the [1-6] bin, which is not very sensitive to the sign of $C_9^\prime$ (contrary to the three large-recoil bin which are in favour of $C_9^\prime<0$). Analysing the [1-6] bin together with low-recoil data yields thus a preference for $C_9^\prime>0$, which is not supported by the $q^2$ dependence of $P_5^\prime$ shown by the three large-recoil bins.
Moreover, if the first low-recoil bin of $B \to K^*\mu^+\mu^-$ (which can be afflicted also by the tail of  charmonium resonances) is removed the preference for $C_9^\prime$ positive of the combined large+low analysis  disappears. An interesting lattice approach to these modes was presented in \cite{lattice}. We asked to the authors of \cite{lattice} to repeat their analysis removing this bin and they found, as we expected, that their best fit point becomes much consistent with zero for $C_9^\prime=+0.4 \pm 0.8$ while $C_9$ remains similar. An extrapolation of their results at low-q$^2$ \cite{stefan} showed also a preference for $C_9^\prime$ negative or zero in this region. In conclusion, the  observed experimental problems at low-recoil in this mode together with the resonance found in this same region in $B^+ \to K^+ \mu^+\mu^-$ precludes from reaching  definite conclusions based on this region contrary to the large-recoil region, until a realistic treatment of errors is implemented at low-recoil.

\subsection{Updated result with experimental correlations}

We have  explored the impact of experimental correlations  between $B \to K^*\mu^+\mu^-$ observables
that were not available at the time of our first analysis using a toy MC technique, as will be described in ref.\cite{nico}. 
An important  correlation, due to the fitting method links the bins of  $A_{FB}$ and $P_2$, and  all other correlations, even if basically negligible, are also included now in \cite{newpaper}. We have repeated our analysis using the same basis as in \cite{thepaper} but including these experimental correlations and we found only a slight reduction of significance from $4.5\sigma$ to $4.2\sigma$. If low-recoil data is also included we obtain 3.5$\sigma$ and 2.7$\sigma$ if only [1-6] bins are considered (see Fig.1).
We have done also an alternative cross check of this result  using $F_L$, instead of $A_{FB}$ always within KMPW parametrization and we obtained the same significances. Finally if the binned ${\cal B}(B \to K^* \mu^+\mu^-)$ (see Fig1) is included in the fit the final significances becomes 4.3$\sigma$ (3.6$\sigma$) and 2.8$\sigma$ respectively.

In conclusion, the result of a full bin analysis using the present LHCb data on $P_i^{(\prime)}$ observables points to a possible explanation of the observed anomaly in terms of a large negative contribution to the coefficient $C_9^{NP} \sim -1.5$, $C_7^{NP}$ negative and small, $C_{10}^{NP}$ small and all other $C_{7,9,10}^\prime$ coefficients also small. 
An important source of information can come from binning the region between 1 to 7\,GeV$^2$. Even if this can naturally imply a reduction in significance due to the smaller statistics and larger experimental error, it can also led to a
 reduction of theoretical uncertainties (impact of $c\bar c$ resonances) a $\chi^2$ improvement  and it might help to clarify the size and sign of $C_9^\prime$, lost in the [1-6] bin analysis. 
 In any case, before drawing any definite conclusion, we should wait for the 3fb$^{-1}$ LHCb data to get a  confirmation of this anomaly or  dismiss it as another statistical fluctuation. 

\medskip
\noindent{\it Acknowledgements}. We thank N. Serra for discussions on the  fitting procedure for correlations.

\end{document}